# Pan-IIT Survey on Online Education - A Report

Survey conducted and reported by[1]:

Achla M. Raina ( IIT Kanpur)
Aditya Nigam ( IIT Mandi)
Amit Gupta ( IIT Delhi)
Amitash Ojha ( IIT Jammu)
Arpan Gupta ( IIT Mandi)
Ashoke Kumar Sarkar ( IIT Jammu)
Badam Singh Kushvah ( IIT Dhanbad)
C. Mahanta ( IIT Guwahati)
Ch. Subrahmanyam ( IIT Hyderabad)
Deepak R ( IIT Palakkad)
Gayathri A (IIT Dharwad)
Devendra Deshmukh ( IIT Indore)
Jose Immanuel R. ( IIT Bhilai)
Kalpesh Haria ( IIT Mandi)
Manoj Gaur ( IIT Jammu)
N S Rajput ( IIT-BHU Varanasi)
Neeraj Goel ( IIT Ropar)
Neeru Chabbra ( IIT Ropar)
Nitin Auluk ( IIT Ropar)

P Seshu ( IIT Dharwad)
Padmanabhan Rajan ( IIT Mandi)
Pravas Ranjan Sahu ( IIT Bhubaneshwar)
Pushpendra Singh ( IIT Jammu)
Rajlakshmi Guha ( IIT Kharagpur)
Rajshekhar Bhat ( IIT Dharwad)
Ramkrishna Pasumarthy ( IIT Madras)
S. R. M. Prasanna ( IIT Dharwad)
S. Senthilvelan ( IIT Guwahati)
Sachin Kore ( IIT Goa)
Sayantan Mandal ( IIT Jammu)
Singh Inderdeep ( IIT Roorkee)
Sridhar Chimalakonda ( IIT Tirupati)
Sridhar Iyer ( IIT Bombay)
Subrata Kumar ( IIT Patna)
Sudipta Sarkar ( IIT Gandhinagar)
Suril Shah ( IIT Jodhpur)
Tom V. Mathew ( IIT Palakkad)
Vaneet Kashyap ( IIT Tirupati)
Vignesh MuthuVijayan ( IIT Madras)

---

[1] The names are in the alphabetical order.

## Background and Motivation

Due to the COVID-19 pandemic that set in in the middle of the semester, most of the IITs (and other institutions from across India and the world) switched to an online mode for completing the semester. Since then, many different institutions from across the world have been discussing ways and means to make online education effective, via various means, such as webinars, talks and short workshops. However, a general observation has been that there is a lack of data on online education, particularly in the Indian context. Anticipating this, we decided to conduct a pan-IIT survey on online education and compiled two sets of questionnaires (one each for faculty members and students) in discussion with the members of the pan-IIT group on Online Pedagogy (https://www.paniit-onlinepedagogy-researchgroup.net/), that has been formed for conducting research on the broad area of online education. The current survey has been steered by IIT Dharwad and IIT Jammu along with the members of the pan-IIT group on online pedagogy.

The survey data was collected from 5th May 2020 to 25th May 2020. We received about 11,890 and 840 responses from students and faculty members, respectively. About 82% and 86% of the respondents were males among the students and faculty surveys, respectively. In the following, we present and discuss the survey responses pertaining to various aspects of online education. An infographic representation of the survey responses may be found in Appendix. We note that the survey conducted was a broad survey, covering various broad aspects of online education. However, a fine-grained survey may be required to obtain an in-depth understanding of specific aspects, such as on how to handle lab courses and carry out assessments.

## Role of Residential Campuses

Right at the outset, we stress that the residential campuses of IITs contribute to the overall growth of an individual in several ways, which are almost impossible to replicate via online education. For instance, the leadership roles in mess, gymkhana, club and other student body activities, where students spend a significant amount of time, cannot be emulated in an online educational setup. Presence of peers is one of the most important aspects of student life. Naturally, about 70% of students were poorly motivated to study, especially due to the absence of peers, according to the current survey. Also, about 38% of the faculty members feel that online education with physical classrooms will become prominent in the future and 27% of the faculty members feel that online education will not be prominent in future. Only 18% of the faculty members feel that online education will take over the classroom education in future. This information indicates that the online education will not (and should not) fully replace the brick-and-mortar classroom education. However, online education will be an important ingredient in education, including undergraduate and graduate curriculum, continuing education programmes and corporate training.

## Responses and Discussion

### Computing and streaming devices

For conducting a decent online education, the students and instructors are required to have a basic set of devices, such as desktops, laptops, tablets or smartphones for preparing and presenting lectures, assignments and lab material for instructors and to attend lectures (synchronous or asynchronous), read books, solve assignments and lab sessions and perform simulations for students. Only about 62% of the students had used laptops/desktops, and about 56% had used smartphones (with about 23% using both). The main reasons why the students did not have access to laptops include the following:

- They did not own one (may be because of economic problems).
- They could not bring it home from the school, as students were given very short notice to vacate the campus and there was no clarity on whether the classes will be held online.

The main challenges faced due to the above include the following:

- Those who did not have a laptop/desktop, faced significant challenges in learning, especially in completing assignments, labs and projects.
- They did not have access to hardcopy of books as they were away from their institute libraries. In addition, they could not download and read the freely available online software and material/references, respectively. Even those who had laptops/desktops, many could not access the material/references as they were not open-source/free.

There were no questions asked to the instructors about the availability of the computing devices, as it is highly likely that they have access to them.

### Connectivity

Seamless internet connectivity is indispensable for online education. Two main options for establishing the connectivity include wired or wireless internet connection.

The survey responses show that only about 20% of the students had stable and good quality internet connection for most of the time. About 80% of the students used mobile data from 3 service providers, namely, Jio, Airtel and Vodafone, which provide 3G/4G services. The connection might have been unstable as mobile data is by nature flickering, especially in remote places with less densely-spaced base stations. In contrast, only 7% of the instructors said they did not have a decent internet connection, as most faculty members stayed in campuses with dedicated networks, and even those who were away would get paid for a telephone/broadband plan.

Hence, not having a good connectivity was one of the major issues among the students.

### Interactive Engagement in Online Classes

About 60% of the instructors felt that lack of real-time interaction is the biggest issue in online education. In the similar lines, about 50% of the students were unable to ask real-time questions due to technology limitation/lack of knowledge to use the options to raise-hand etc. About 50% of the faculty members want to increase the real-time interaction in the next online class they teach. About 58% of the students felt that online education can be improved making them interactive using technology.  It is interesting to note that more students feel that the online education can be improved by providing better quality audio/video (74% of the students) and  giving written/typed notes (61% of the students), than by having interactive classes (only 58% of the students).

### Mode of Instruction

The instruction was carried out both in the synchronous and asynchronous modes. About 45% of the faculty members conducted classes in a synchronous mode (mainly via Zoom and Google Meet) and about 42% of the instructors conducted classes in an asynchronous mode. In the synchronous mode, the instructors asked for students' feedback explicitly, as it was difficult to decipher it through non-verbal feedback of the students. In the asynchronous mode, about 26% of the instructors recorded handwritten slides/notes via  mobile phones.

 More than 55% of the students had online classes for less than 5 hours in the synchronous mode. Only about 40% of the students were able to attend more than 50% of online classes without any difficulty.  The major difficulties faced  were unclear audio/video (61%) and annoying video buffering (56%). Clearly, this is due to the fact that the majority of the students had difficulty in accessing good quality internet data. In the asynchronous mode, about 45% of the students viewed recorded lectures for less than 5 hours. Only 25% of the students were able to view recorded classes without difficulty. About 50% of the students were unable to download the material shared with them.   Other problems included unclear audio and videos and unclear explanation by the instructors.

 About 47%, 14% and 40% of the instructors prefer online education only in synchronous, only in asynchronous and both the modes, respectively. On the other hand, 20%, 28% and 55% of the students prefer online education only in synchronous, only in asynchronous and both the modes, respectively.  This may be because, from the instructor-perspective, the  preparation time (and time to achieve perfection, if obsessed about it) for a synchronous class is shorter than that for an asynchronous class and the synchronous mode is to be more interactive compared to the asynchronous mode.  However, from the student-perspective, in a synchronous class, they do not get to exploit the self-paced learning aspect of the online class. Moreover, in case of poor internet connection, they may not even be able to decipher what the instructor is speaking  in the first place, let alone understand it.

*Preparedness*

About 74% of the faculty members did not have prior experience of conducting fully online courses. Other 26% faculty members had given courses like NPTEL/SWAYAM. However, About 88% of the faculty members had medium to high preparedness for conducting classes online of some form. This may be because even in the in-campus teaching, except lecture delivery and exams, other aspects (assignments, sharing material etc.) are done online. Moreover, it is interesting to note that about 80% of the instructors' satisfaction level was medium to very high. However, they want to improve by making online education more interactive.

On the other hand, more than 60% of the students were under-prepared for the online classes. Due to the short-notice and uncertainty, they had not carried books, notes, laptops etc. The books were not available online for easy access. Due to this, about 46% of the students had difficulty in completing the assignments. About 18% of the students had difficulty in completing assignments as they did not have the required software installed in their computer.

*Assessment*

About 46% of the instructors did not conduct any assessment. 27% of them gave open-book/open-internet exams and 12% conducted viva-voce exams. Others conducted online proctored exams, term-paper presentations etc. When asked about the suggestions on assessment, about 59% of the instructors felt that the assessment should include multiple components, such as quizzes, assignments, course projects and classroom questions. About 32% of the instructors suggested conducting limited-time open-book/internet exams and viva-voce exams, only about 28% suggested online proctored traditional exams. More or less a similar trend may be observed in the students' responses too. About 61% felt that no exams should be conducted. 41% and 17% of the students respectively felt that open-book/internet exams and viva-voce exams should be conducted. Only 11% suggested conducting remotely proctored exams. It is interesting to note that both the instructors and the students indicate that we should move away from the traditional assessment techniques, such as conducting proctored exams. Hence, challenges remain in finding a reliable, robust and transparent method of conducting assessment. New methods need to be proposed, experimented and improved.

*Role of Teaching Assistants (TAs)*

About 88% of the students did not interact with the TAs. In the similar lines, 36% of the instructors felt that the contribution of the TAs has been lower during the online classes. These numbers clearly show that the TAs were under-utilized. When trained properly, they may help in taking the load of post-processing of content that may be required for asynchronous classes (recall that only 14% of the instructors preferred asynchronous mode, but 28% of the students wanted classes in asynchronous mode).

### Laboratory Classes

About 68% of the classes conducted by the instructors did not have any lab component. 10% of them who had software-based labs conducted them. However, about 3% of them could not conduct the software-based labs due to unavailability of off-campus licenses. About 11% of the courses required physical labs and hence they could not be conducted. Interestingly, 2% of the instructors transformed their physical lab to a simulation-based lab and conducted them. When it comes to projects, only 27% of the students had no difficulty in executing them. The difficulties were the following: 28% of the students had projects that required physical lab access, 24% required software, which was not available, and 29% could not access reference material that was required for their projects. These responses show that the laboratories could not be conducted online, especially when it involves physical equipment, in almost all the major core disciplines such as aerospace, civil, electrical and mechanical engineering. Finally, among about 1300 PhD students who responded, about 54% could not carry out their lab experiments and research online at all.

### Challenges in Working from Home

The main challenge in working from home for the instructors was lack of ergonomically designed office spaces (52%). Other problems faced were disturbance due to kids seeking constant attention (34%) and noise from TV, neighbours etc. (28%). Similarly, lack of study rooms, noise from family members, TV and neighbours/street-vendors were the main challenges for the students. Due to these, about 72% of the students had lower-productivity at home and about 70% of the students had low to medium motivation levels to study at home.

### Correlation between Various Factors

In this part, we mention insights from an initial correlation analysis that was conducted.

- Download speed positively correlates with experience of male students/ UG students and motivation level of PhD students. Upload speed does not seem to have any positive correlation with anything except having negative correlation with experience of male participants.
- Recorded classes motivate UG students in general. But it has a negative correlation with the experience of male students but positive correlation with the experience of female students.
- Live classes do not motivate students in general but have high positive correlation with the experience of students.
- Heavy loads of assignments do not motivate boys but have no impact on females. With boys' motivation, it has a negative correlation. Heavy loads of assignments do not significantly impact girls' motivation.
- Heavy load of assignments has a highly positive correlation with experience of girls. Low load of assignment has high correlation with boys motivation but does not correlate with motivation of girls but it has positive correlation with the experience of both boys and girls.

## Conclusion

The online education has several advantages, such as self-paced learning, any-time and any-place learning. However, efforts are needed to improve online education. The starting point is to improve the availability of computing/streaming devices and enabling good internet connectivity among the students. The next step is to prepare decent quality content and deliver it to students – here majority of the faculty members and students prefer a combination of synchronous and asynchronous modes. Lack of real-time interaction is also an issue both among instructors and faculty members. Unavailability of textbooks, reference material and software are also big challenges as many students could not complete their assignments, labs and projects due to this. Lack of a campus-like environment at home is also a serious issue in online education. From the survey, it is clear that addressing the above issues is critical for making online education a success. Overall, it therefore appears challenging to visualise online education replacing classroom education for technical education at this point in time. However, it could certainly supplement regular classroom education for our main offerings, such as BTech, MTech and MS/PhD programs which will still have to continue in regular mode. Nevertheless, with improvements in technology and its accessibility, high quality online education is expected to become pervasive.

# Pan-IIT Faculty Survey on Online Education

---

## DEMOGRAPHICS

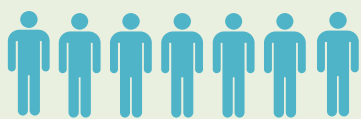
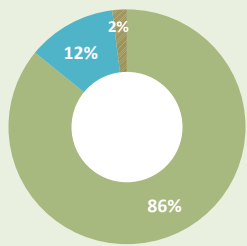
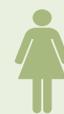

2%
12%
86%

**Males**
About 86% of the respondents are males

**Females**
About 12% of the respondents are females

**Observations:**
About 7 male faculty members responded per 1 female faculty member. 2% of the respondents preferred to not reveal their gender.

# Designation

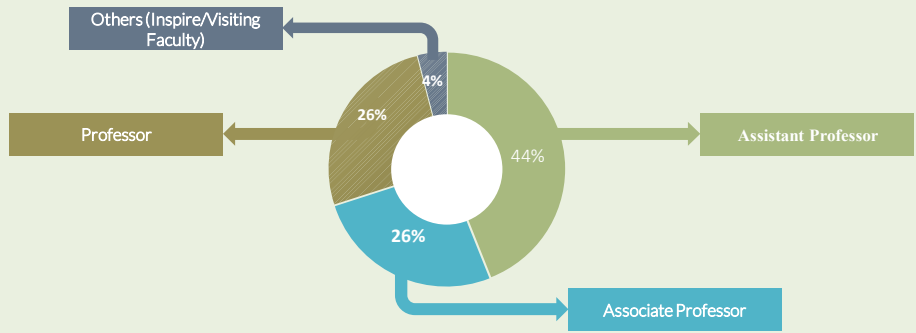

**Observations:**

About 70% of the respondents were Assistant or Associate Professors.

# Prior Experience of Online Education

**Observations:**

About 74% of the faculty members did not have prior experience of conducting fully online courses. Other 26% faculty members have given courses like NPTEL/SWAYAM.

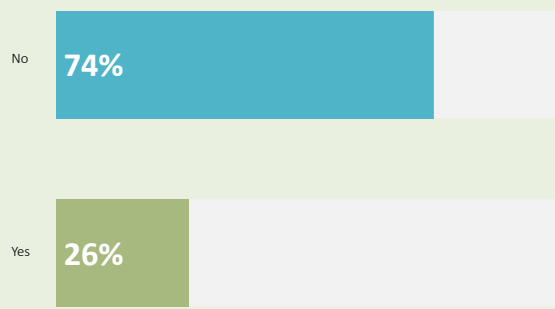

# Preparedness for the Online Classes

**Observations:**

About 88% of the faculty members had medium to high preparedness for conducting classes online. This may be because even in the in-campus teaching, except lecture delivery and exams, other aspects (assignments, sharing material etc.) are done online.

- High: **54%**
- Medium: **34%**
- Low: **12%**

---

# PROGRAM OF STUDY

**Observations:**

Major fractions of instructors conducted online classes for first year masters students and final year undergraduate students.

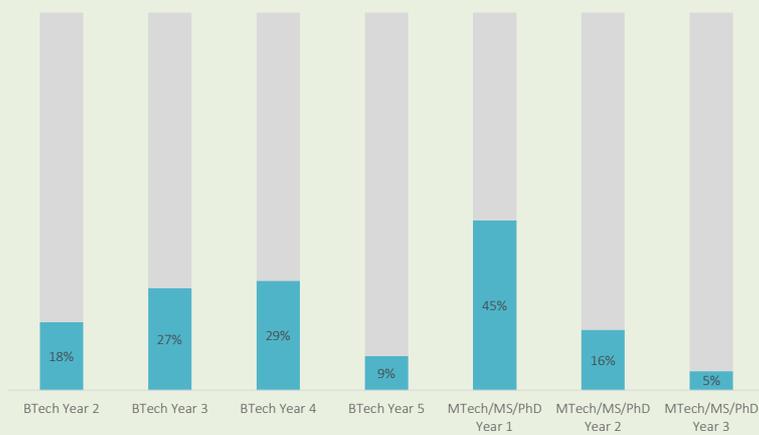

| BTech Year 2 | BTech Year 3 | BTech Year 4 | BTech Year 5 | MTech/MS/PhD Year 1 | MTech/MS/PhD Year 2 | MTech/MS/PhD Year 3 |
|---|---|---|---|---|---|---|
| 18% | 27% | 29% | 9% | 45% | 16% | 5% |

## Course Participants

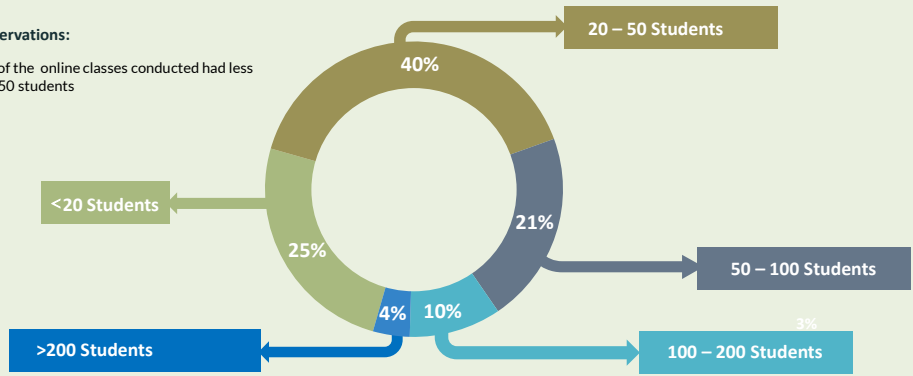

**Observations:**
65% of the online classes conducted had less than 50 students

- 20 – 50 Students: 40%
- <20 Students: 25%
- 50 – 100 Students: 21%
- 100 – 200 Students: 10%
- >200 Students: 4%
- 3%

## Time-Table for Online Classes

**Observations:**
Majority of the faculty members re-scheduled their classes from regular slots when they switched to the online mode.

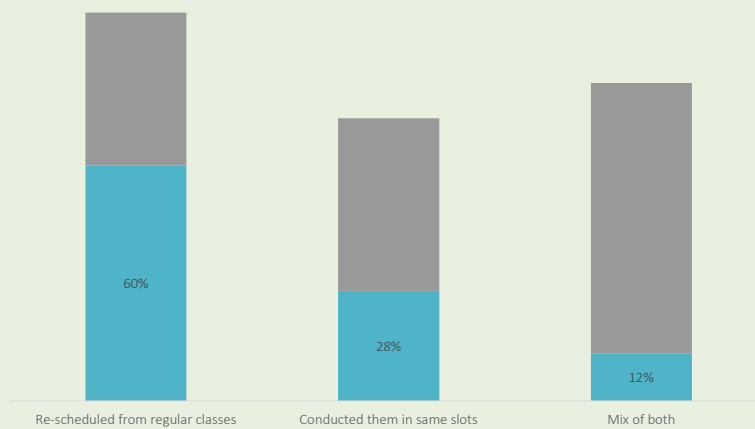

- Re-scheduled from regular classes: 60%
- Conducted them in same slots: 28%
- Mix of both: 12%

# Core or Elective Class?

**Observations:**
Among the conducted courses, about 55% were core and 55% were elective courses. Note that some of the courses may be core for some and elective for other students.

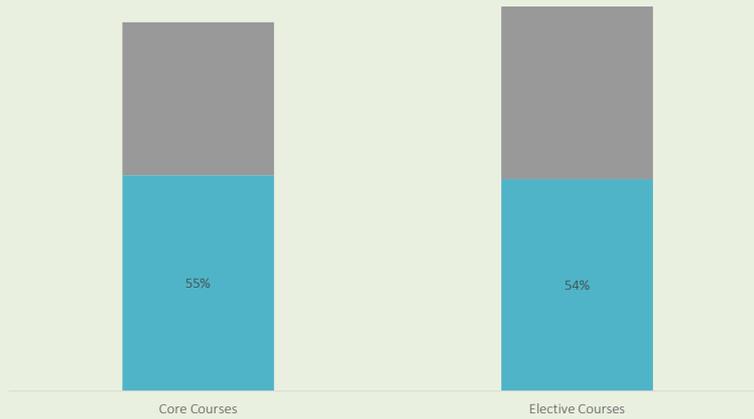

# Minimum Data/Day Required for Online Classes

**Observations:**
About 30% instructors felt that 1GB/day data and 45% felt that more than 2GB/day data is required for offering decent online classes (instructor-side). Others have not checked.

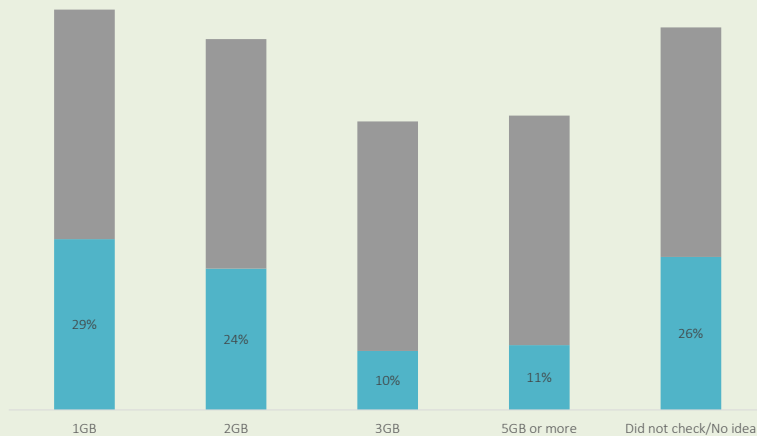

## Minimum Data Speed for Required for Online Classes

**Observations:**
About 40% of the instructors feel that about 5Mbps speed is required for conducting decent online class (instructor-side).

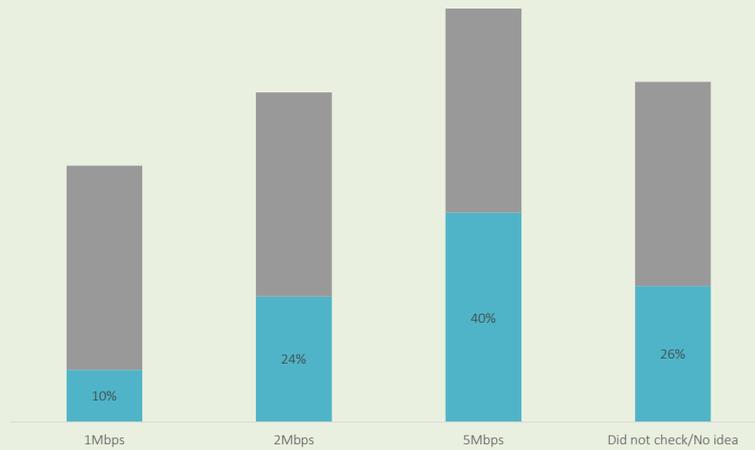

| 1Mbps | 2Mbps | 5Mbps | Did not check/No idea |
|---|---|---|---|
| 10% | 24% | 40% | 26% |

---

## Quality of Internet Connection

**Observations:**
Majority of the faculty members had decent internet connection.

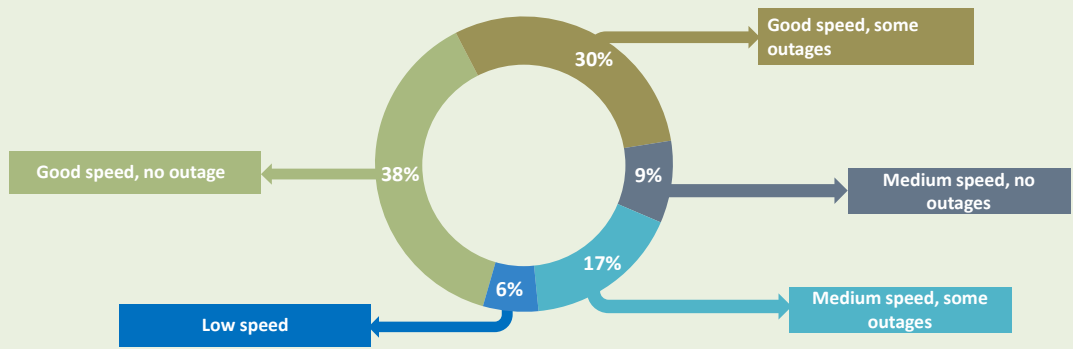

- Good speed, no outage: 38%
- Good speed, some outages: 30%
- Medium speed, no outages: 9%
- Medium speed, some outages: 17%
- Low speed: 6%

## How Many Weeks of Online Classes?

**Observations:**

About 70% of the classes were held for more than 3 weeks.

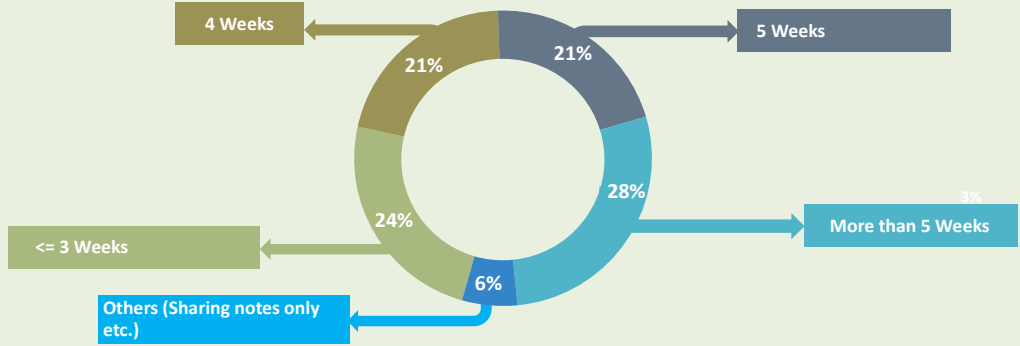

- 4 Weeks — 21%
- 5 Weeks — 21%
- More than 5 Weeks — 28%
- Others (Sharing notes only etc.) — 6%
- <= 3 Weeks — 24%

## Hours/Week

**Observations:**

About 43% of the classes had instruction for more than 3 hours/week.

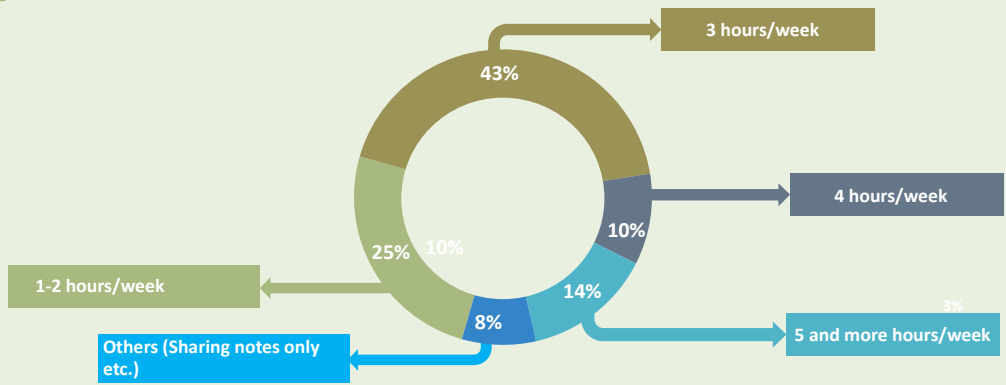

- 3 hours/week — 43%
- 4 hours/week — 10%
- 5 and more hours/week — 14%
- Others (Sharing notes only etc.) — 8%
- 1-2 hours/week — 25%

## Mode of Delivery

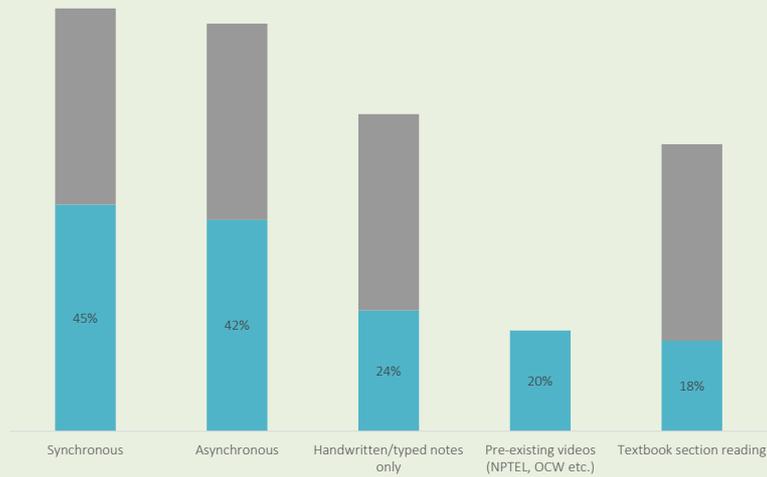

**Observations:**
About 45% of the instructors conducted classes in synchronous mode.

| Synchronous | Asynchronous | Handwritten/typed notes only | Pre-existing videos (NPTEL, OCW etc.) | Textbook section reading |
|---|---|---|---|---|
| 45% | 42% | 24% | 20% | 18% |

## Tools Used for Synchronous Mode

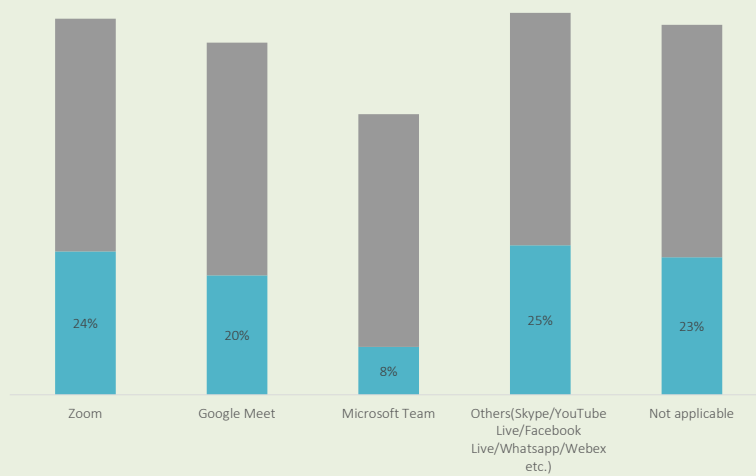

**Observations:**
Zoom and Google Meet are the most widely used tools for synchronous delivery of lectures.

| Zoom | Google Meet | Microsoft Team | Others (Skype/YouTube Live/Facebook Live/Whatsapp/Webex etc.) | Not applicable |
|---|---|---|---|---|
| 24% | 20% | 8% | 25% | 23% |

## Tools Used for Asynchronous Mode

**Observations:**

About 26% of the instructors recorded handwritten slides/notes via mobile phones. Looks like there is no preference for a tool over others, as many used various different tools.

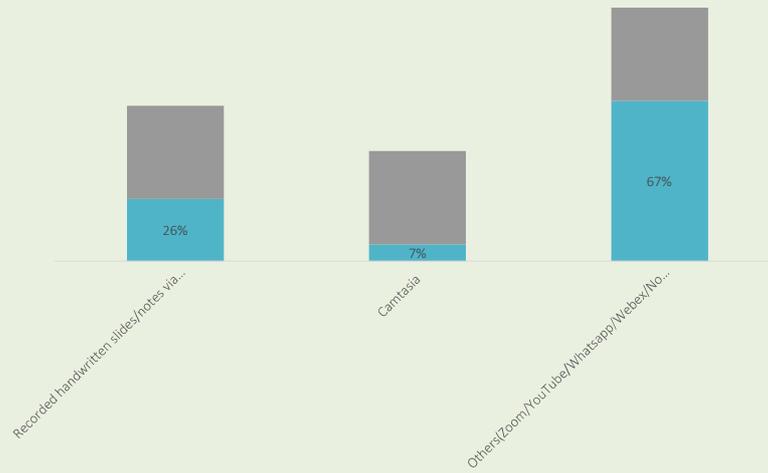

## Doubt Clarification

**Observations:**

Only 40% of the instructors were able to clarify doubts in the synchronous mode.

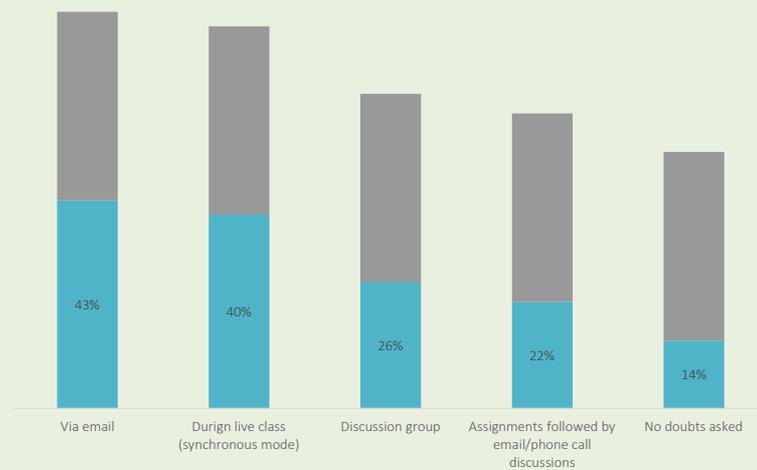

# Gauging Student Feedback

**Observations:**
Live videos from students were not really useful for gauging students' feedback during live classes.

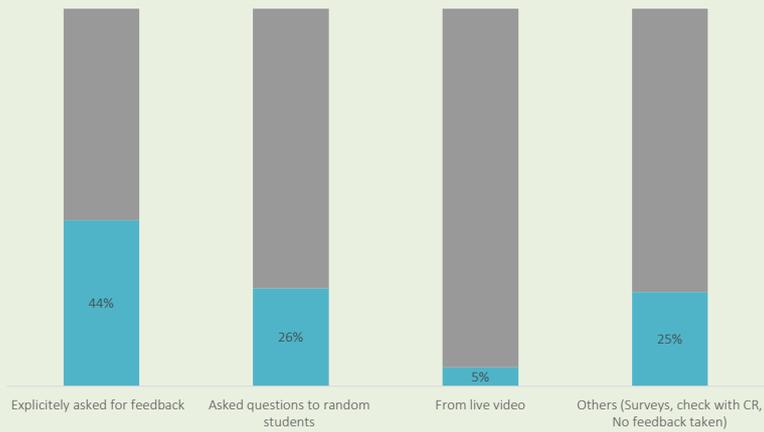

- Explicitely asked for feedback: 44%
- Asked questions to random students: 26%
- From live video: 5%
- Others (Surveys, check with CR, No feedback taken): 25%

# Material Shared Via?

**Observations:**
Sharing material via email is the most widely used method.

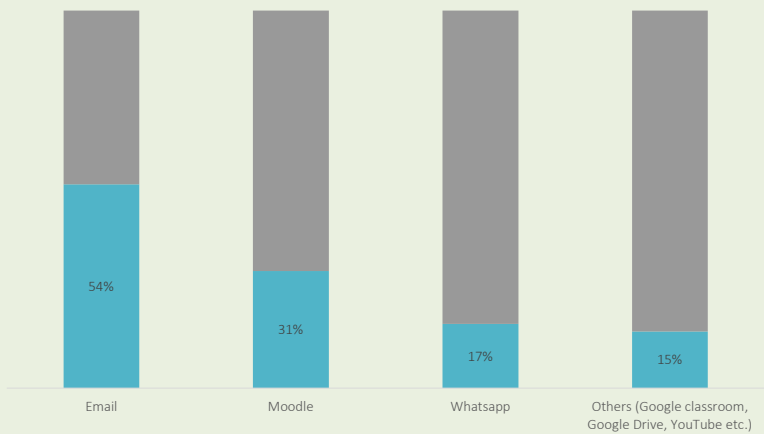

- Email: 54%
- Moodle: 31%
- Whatsapp: 17%
- Others (Google classroom, Google Drive, YouTube etc.): 15%

## How Did You Emulate Black/White Board Work?

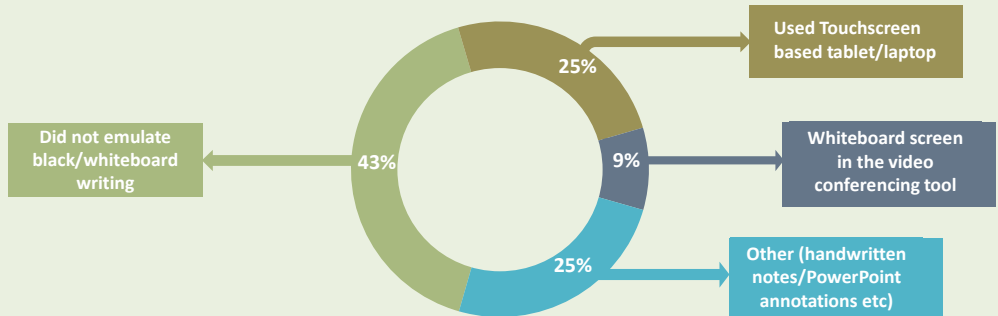

- 25% Used Touchscreen based tablet/laptop
- 9% Whiteboard screen in the video conferencing tool
- 25% Other (handwritten notes/PowerPoint annotations etc)
- 43% Did not emulate black/whiteboard writing

**Observations:**
About 43% of instructors did not emulate black/whiteboard writing.

---

## How Did You Conduct Assessment?

**Observations:**
Most instructors did not conduct any assessment.

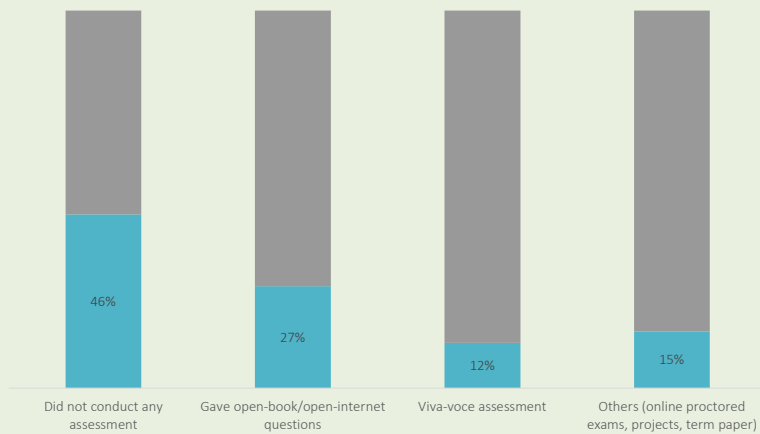

- Did not conduct any assessment: 46%
- Gave open-book/open-internet questions: 27%
- Viva-voce assessment: 12%
- Others (online proctored exams, projects, term paper): 15%

## How Did You Conduct Labs?

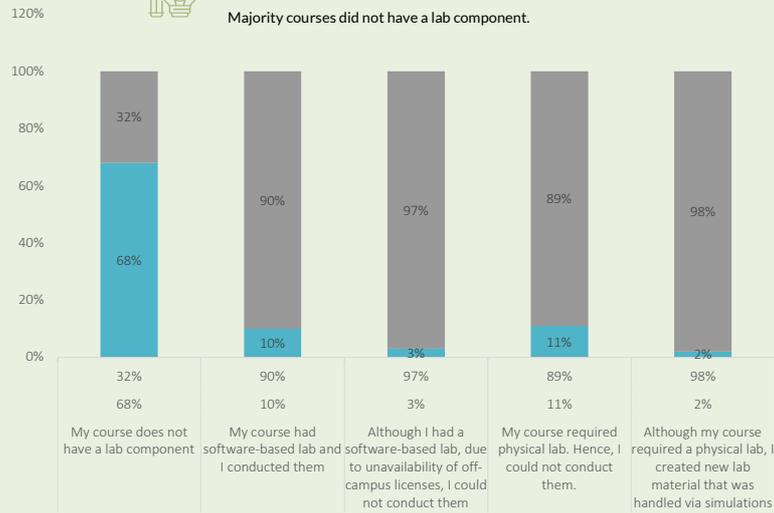

**Observations:**
Majority courses did not have a lab component.

## Issues in Working from Home

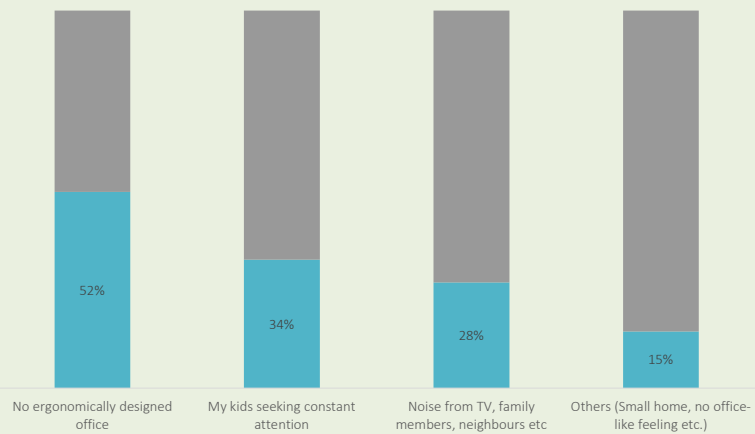

**Observations:**
Lack of office setup was the main shortcoming when working from home.

# Challenges for Conducting Online Classes Effectively

**Observations:**
Lack of interaction with students is the biggest challenge in conducting online classes effectively.

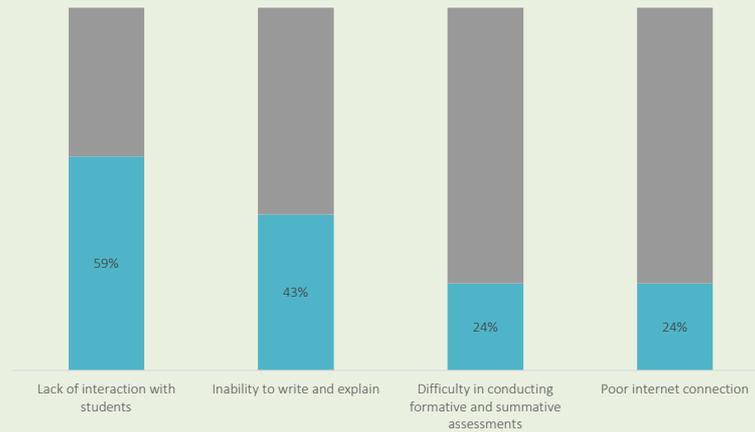

- Lack of interaction with students: 59%
- Inability to write and explain: 43%
- Difficulty in conducting formative and summative assessments: 24%
- Poor internet connection: 24%

# % of Students Actively Participating in Classes

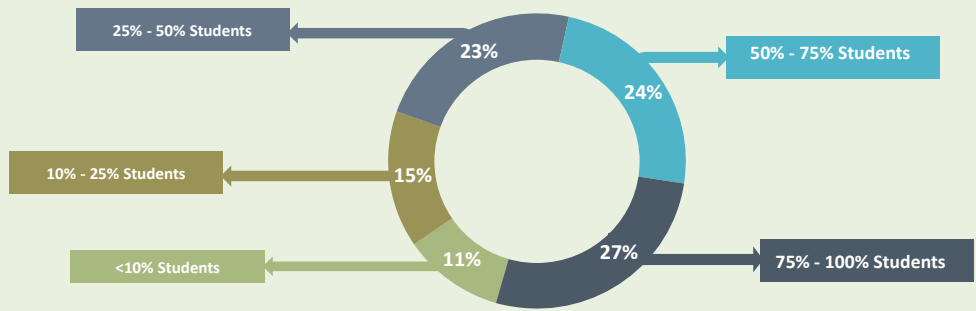

- 25% - 50% Students: 23%
- 50% - 75% Students: 24%
- 10% - 25% Students: 15%
- 75% - 100% Students: 27%
- <10% Students: 11%

**Observations:**
About 50% instructors felt that more than 50% of the students have actively participated in online classes.

## Student Participation in Online Classes compared to Physical Classes

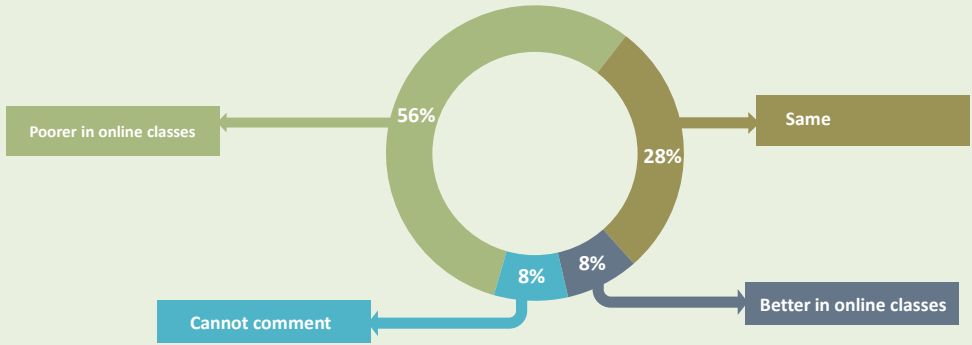

- Poorer in online classes: 56%
- Same: 28%
- Cannot comment: 8%
- Better in online classes: 8%

**Observations:**

More than 50% instructors felt that participation of students were poorer in online classes. This may be partially because of lack of personal interaction.

---

## Role of TAs in Online Classes Compared to Physical Classes

**Observations:**

Majority instructors felt that contribution from TAs has been lower during online classes compared to in-campus classes.

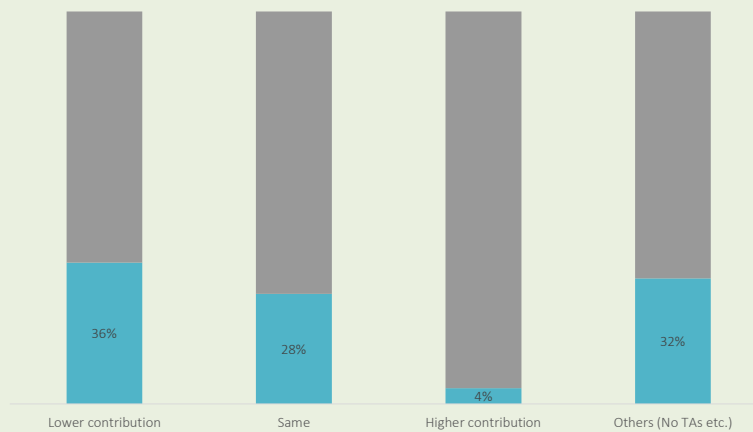

| Lower contribution | Same | Higher contribution | Others (No TAs etc.) |
|---|---|---|---|
| 36% | 28% | 4% | 32% |

## Preparation Time Required for Online Classes Compared to Regular Classes

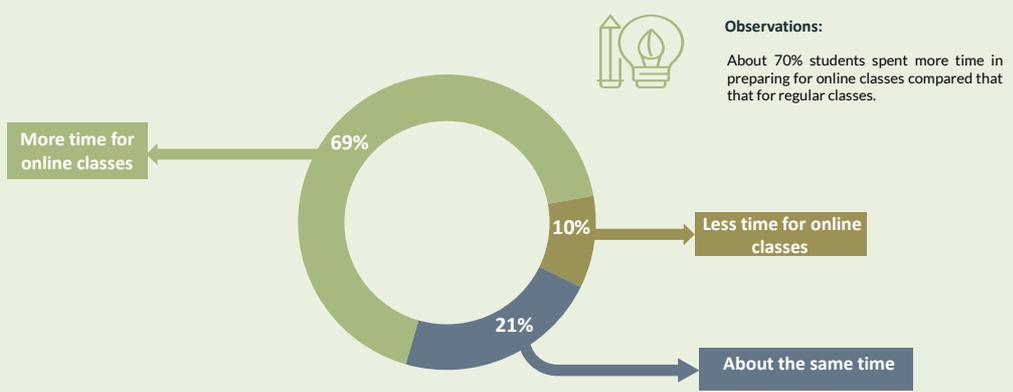

**Observations:**

About 70% students spent more time in preparing for online classes compared that that for regular classes.

- 69% More time for online classes
- 10% Less time for online classes
- 21% About the same time

## Satisfaction Level with Online Classes Conducted

**Observations:**

About 80% of the instructors' satisfaction level was medium to very high.

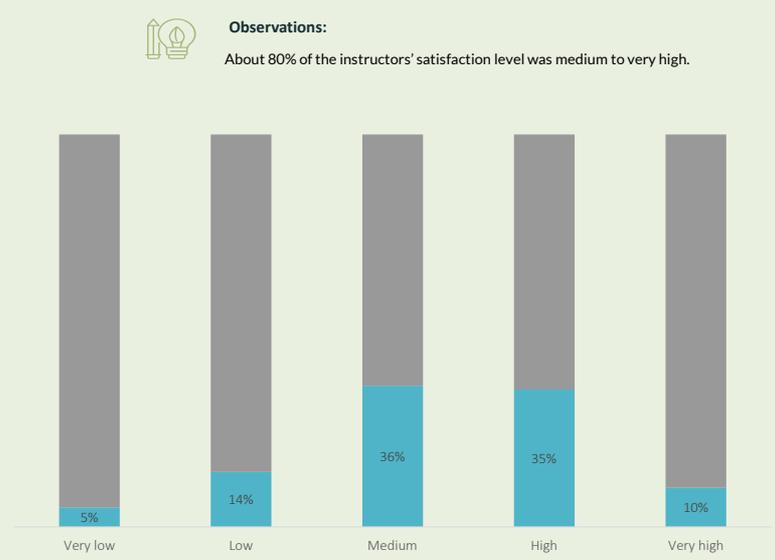

| Very low | Low | Medium | High | Very high |
|---|---|---|---|---|
| 5% | 14% | 36% | 35% | 10% |

## What changes you plan to make on your next online class?

**Observations:**
Majority of the instructors hope to increase the live interaction with students.

- 48% — Try to increase live interaction
- 17% — Will teach only essential material live and share other material offline.
- 11% — Will embed demos, visuals, online program running
- 26% — Others (modular videos etc.)

## Advantages of Online Education over Brick-and-Mortar Classes

**Observations:**
Majority of the instructors feel self-paced learning is the greatest advantage of online education.

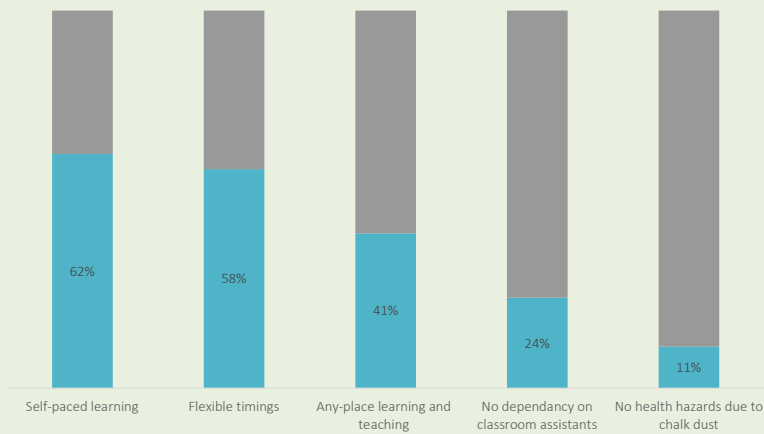

- Self-paced learning: 62%
- Flexible timings: 58%
- Any-place learning and teaching: 41%
- No dependancy on classroom assistants: 24%
- No health hazards due to chalk dust: 11%

## Disadvantages of Online Education over Brick-and-Mortar Classes

**Observations:**

Lack of personal interaction and difficulty in gauging students' non-verbal feedback seems to be the biggest disadvantage of online education.

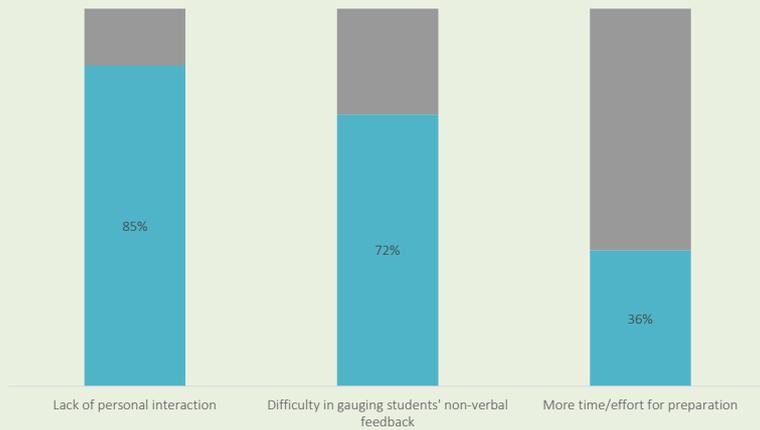

- Lack of personal interaction: 85%
- Difficulty in gauging students' non-verbal feedback: 72%
- More time/effort for preparation: 36%

## Will Online Education be the Future of Education?

**Observations:**

Only 18% of the instructors feel that pure online education will be the future of education.

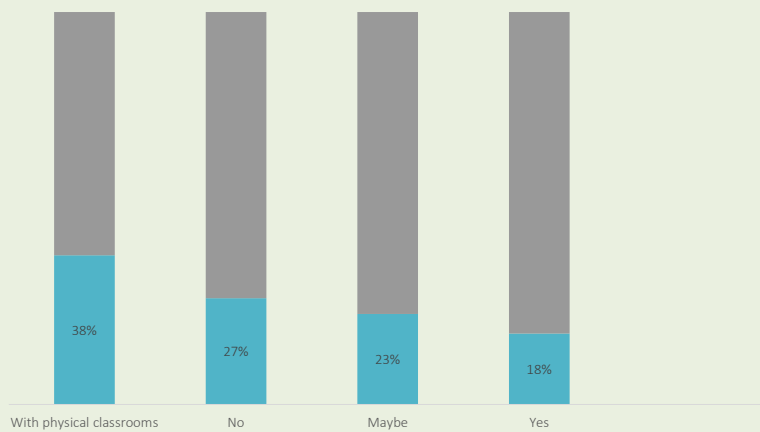

- With physical classrooms: 38%
- No: 27%
- Maybe: 23%
- Yes: 18%

## ONLY DELIVERY MODE PREFERENCE

**Synchronous Mode** 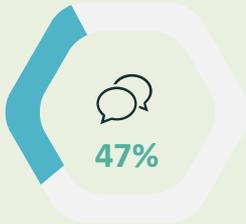
**47%**

**Only Asynchronous Mode** 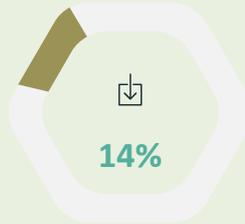
**14%**

**A Combination of Both** 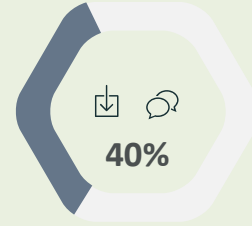
**40%**

**Observations:**

Majority of instructors prefer synchronous mode of delivery.

---

## What policy changes we need to embrace online education in a large scale?

**Observations:**

For moving towards large scale online education, majority of the instructors feel that we must start with some courses having some online delivery components.

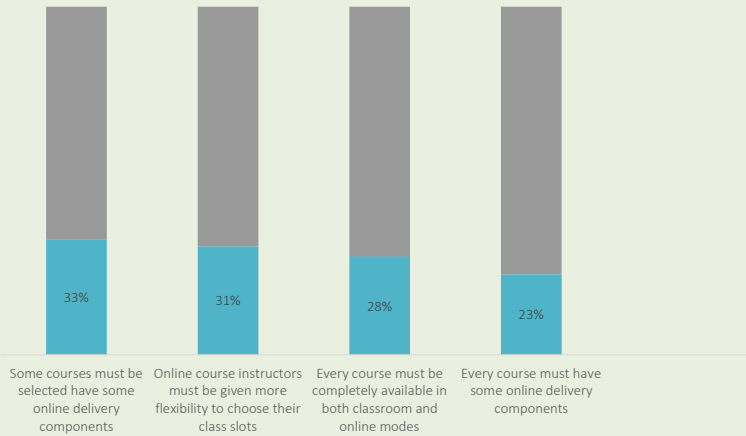

- 33% — Some courses must be selected have some online delivery components
- 31% — Online course instructors must be given more flexibility to choose their class slots
- 28% — Every course must be completely available in both classroom and online modes
- 23% — Every course must have some online delivery components

# Suggestions for Conducting Online Exams

**Observations:**
About 60% of instructors feel that we must have multiple components, such as assignments, quizzes, course projects and classroom questions, in evaluating students in an online course.

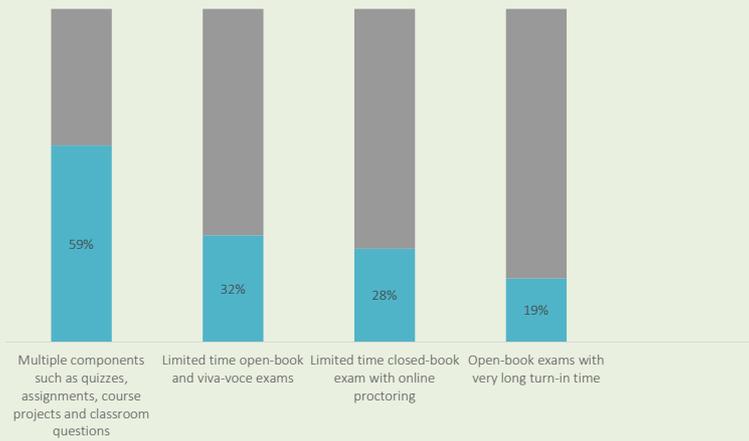

| | | | |
|---|---|---|---|
| 59% | 32% | 28% | 19% |
| Multiple components such as quizzes, assignments, course projects and classroom questions | Limited time open-book and viva-voce exams | Limited time closed-book exam with online proctoring | Open-book exams with very long turn-in time |

# Pan-IIT Student Survey on Online Education

---

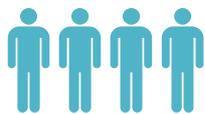 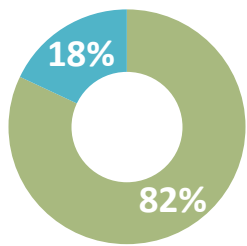 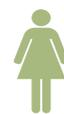

**18%**

**82%**

**Males**

About 82% of the respondents are males

**Females**

About 18% of the respondents are females

**Observations:**

**About 4 male students responded per 1 female student. This is roughly the same as the male-to-female ratio in admissions at IITs.**

# BROAD CATEGORY OF COURSES

**Observations:**

80% of the students are from engineering background. Among them, some are in multiple categories.

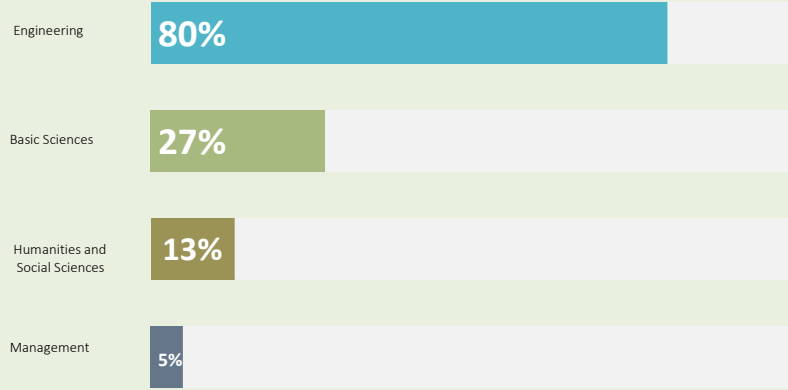

- Engineering: **80%**
- Basic Sciences: **27%**
- Humanities and Social Sciences: **13%**
- Management: **5%**

---

# PROGRAM OF STUDY

**Observations:**

Among those who responded, 68% are from undergraduate studies.

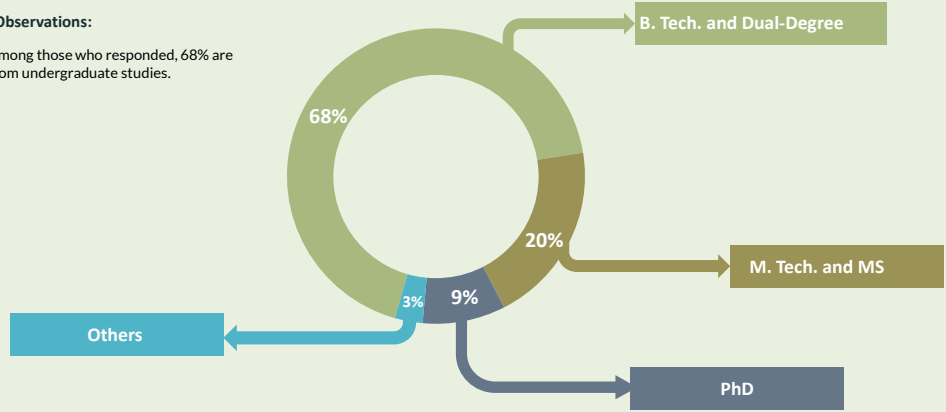

- B. Tech. and Dual-Degree: 68%
- M. Tech. and MS: 20%
- PhD: 9%
- Others: 3%

# YEAR OF STUDY

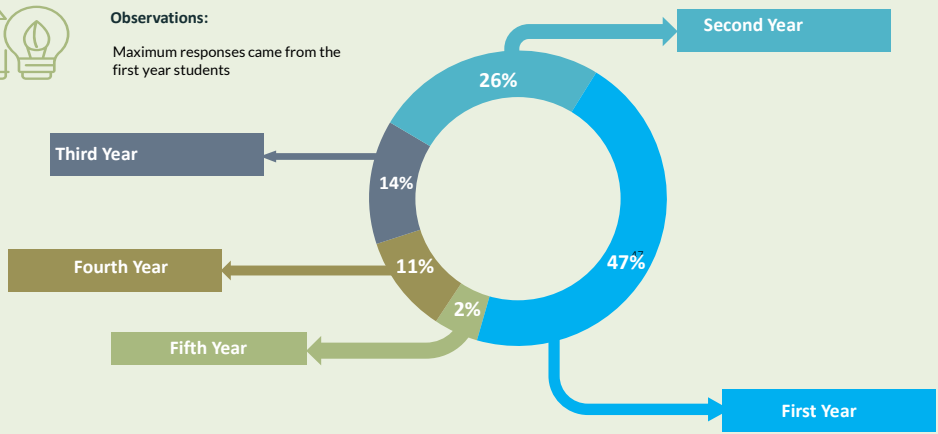

**Observations:**
Maximum responses came from the first year students

- Second Year: 26%
- Third Year: 14%
- Fourth Year: 11%
- Fifth Year: 2%
- First Year: 47%

# SGPA

**Observations:**
About 80% of the students who have responded have more than 7.0 SGPA (above average performance).

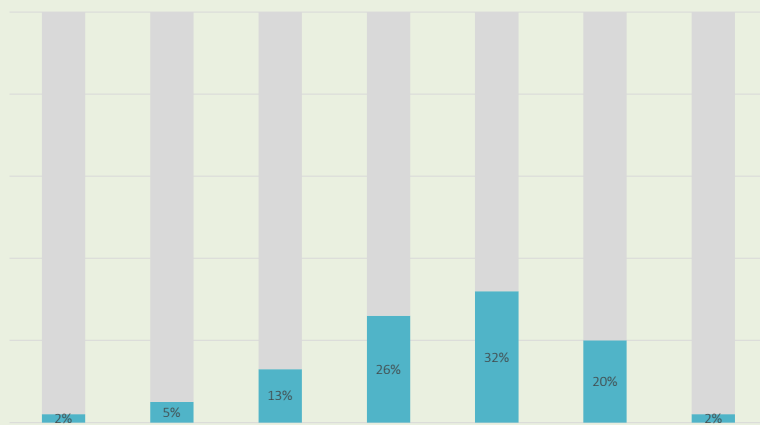

2% | 5% | 13% | 26% | 32% | 20% | 2%

# CGPA

**Observations:**
About 80% of the students who have responded have more than 7.0 CGPA (above average performance).

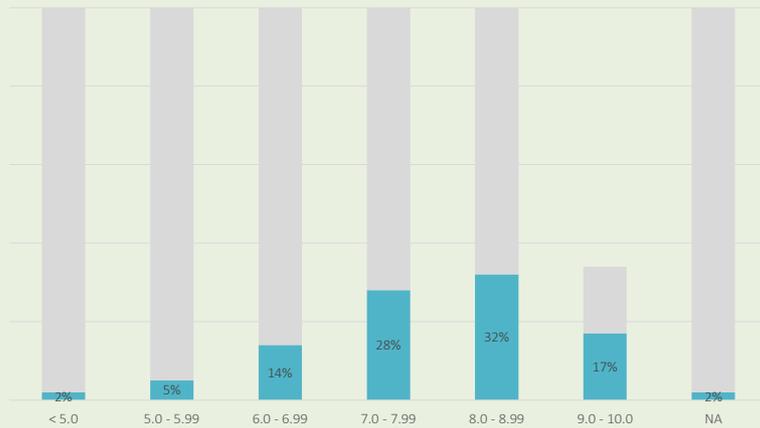

| < 5.0 | 5.0 - 5.99 | 6.0 - 6.99 | 7.0 - 7.99 | 8.0 - 8.99 | 9.0 - 10.0 | NA |
|---|---|---|---|---|---|---|
| 2% | 5% | 14% | 28% | 32% | 17% | 2% |

# % SYLLABUS COVERED IN ONLINE MODE

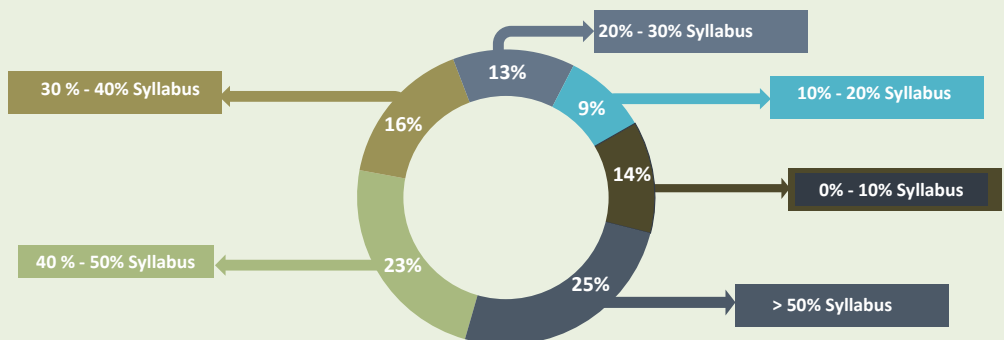

- 20% - 30% Syllabus: 13%
- 10% - 20% Syllabus: 9%
- 0% - 10% Syllabus: 14%
- > 50% Syllabus: 25%
- 40% - 50% Syllabus: 23%
- 30% - 40% Syllabus: 16%

**Observations:**
More than 40% of the syllabus was covered by online mode for about 50% of the students.

## PREPARATION LEVEL FOR ONLINE CLASSES

- Medium: 27%
- Under-prepared: 62%
- Well-Prepared: 11%

PhD

**Observations:**

More than 60% of the students were under-prepared for the online classes. Due to uncertainty, they had not carried books, notes, laptops etc. The books were not available online for easy access

---

## INTERNET SPEEDS

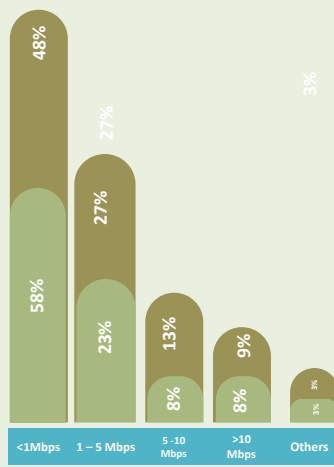

| Speed | Upload | Download |
|---|---|---|
| <1Mbps | 48% | 58% |
| 1 – 5 Mbps | 27% | 23% |
| 5-10 Mbps | 8% | 13% |
| >10 Mbps | 8% | 9% |
| Others | 3% | 3% |

**Upload speed**

Upload speed was not sufficient for good quality audio and video conversation for more than 50% of the students

**Download speed**

Download speed was not sufficient for good quality audio and video conversation for more than 50% of the students

# STABILITY/QUALITY OF INTERNET CONNECTION

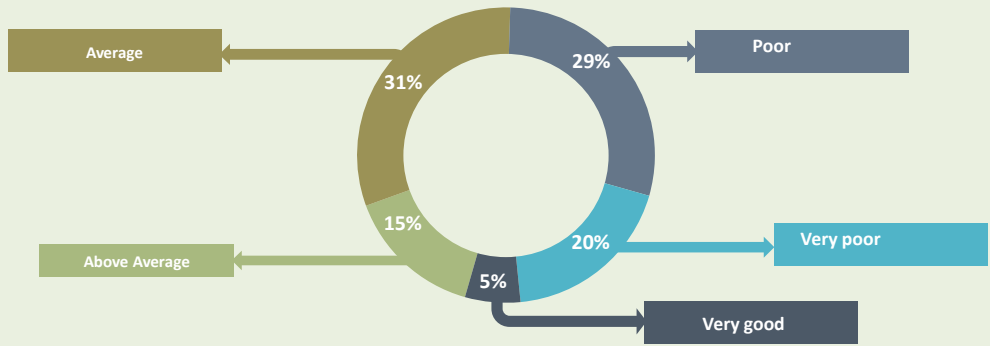

- Average 31%
- Poor 29%
- Very poor 20%
- Very good 5%
- Above Average 15%

**Observations:**

Only about 20% of the students have stable, good quality Internet connection for most of the time.

---

# INTERNET SERVICE PROVIDERS

**Observations:**

Most people use mobile data from Jio and Airtel.

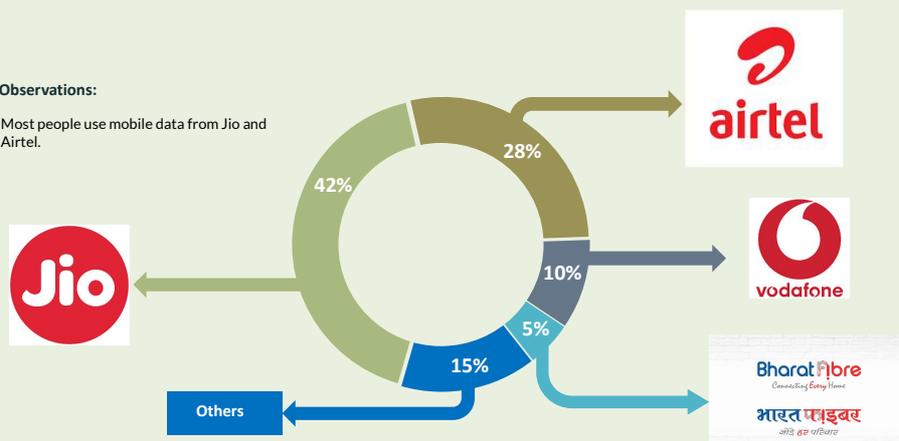

- Airtel 28%
- Jio 42%
- Vodafone 10%
- Bharat Fibre 5%
- Others 15%

## POWER OUTAGE DURATION/DAY

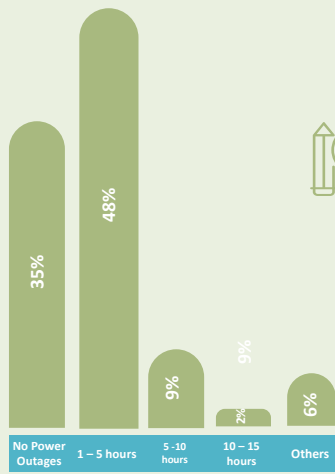

35% — No Power Outages
48% — 1 – 5 hours
9% — 5 -10 hours
2% — 10 – 15 hours
9% (label)
6% — Others

**Observations:**

More than 50% of the students have some form of power outages. It is severe among about 10% of the students.

The major problems faced due to power cuts are Internet outages. Other problems include inability to charge phones and laptops, and not getting sound sleep.

---

## Devices Used

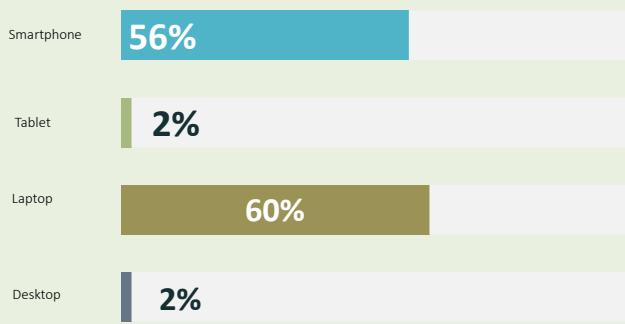

Smartphone — 56%
Tablet — 2%
Laptop — 60%
Desktop — 2%

**Observations:**

This shows that about 60% of the students had access to laptops. Given that most of the engineering education requires laptop/desktop, remaining 40% of the students seem to have problem.

# Tools Used for Online Classes

**Observations:**
Variety of different tools were used. About 41% of the instructors shared hand-written/typed notes.

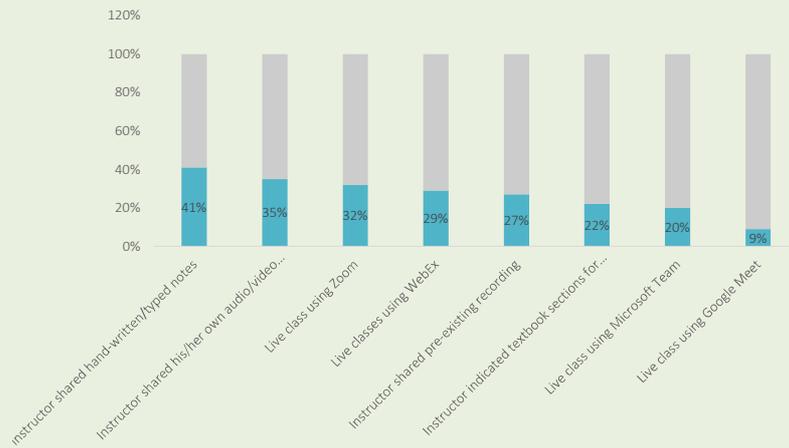

| Tool | % |
|---|---|
| Instructor shared hand-written/typed notes | 41% |
| Instructor shared his/her own audio/video... | 35% |
| Live class using Zoom | 32% |
| Live classes using WebEx | 29% |
| Instructor shared pre-existing recording | 27% |
| Instructor indicated textbook sections for... | 22% |
| Live class using Microsoft Team | 20% |
| Live class using Google Meet | 9% |

# Motivation Levels

**Observations:**
About 70% of the students had low to medium motivation levels to study at home.

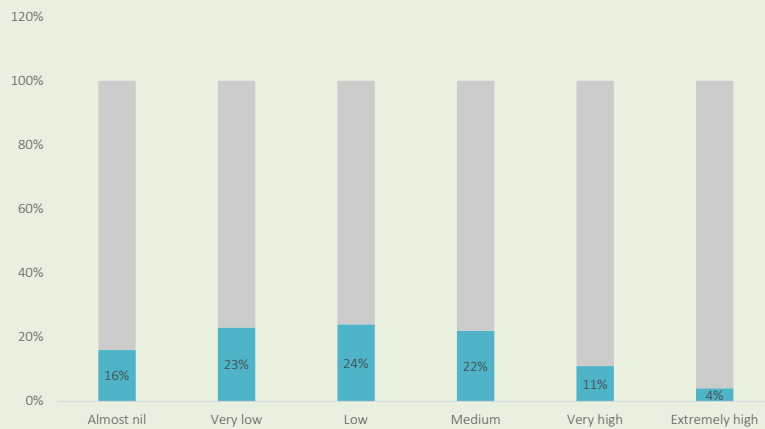

| Level | % |
|---|---|
| Almost nil | 16% |
| Very low | 23% |
| Low | 24% |
| Medium | 22% |
| Very high | 11% |
| Extremely high | 4% |

# #Hours – Synchronous Mode

**Observations:**
More than 55% of the students had online classes for less than 5 hours.

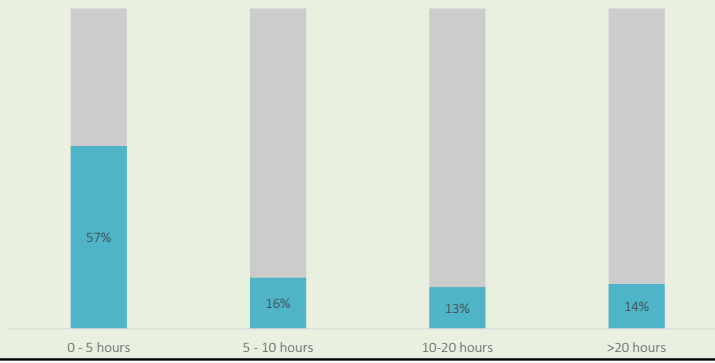

# % Classes Attended w/o Difficulty - Synchronous Mode

**Observations:**
Only about 40% of the students were able to attend more than 50% of online classes without any difficulty.

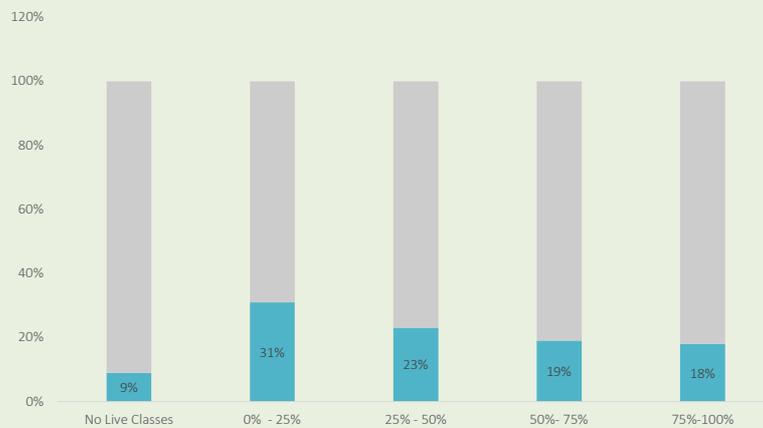

## Nature of Difficulty Faced - Synchronous Mode

**Observations:**
For about 60% of the students, audio/video was not clear.

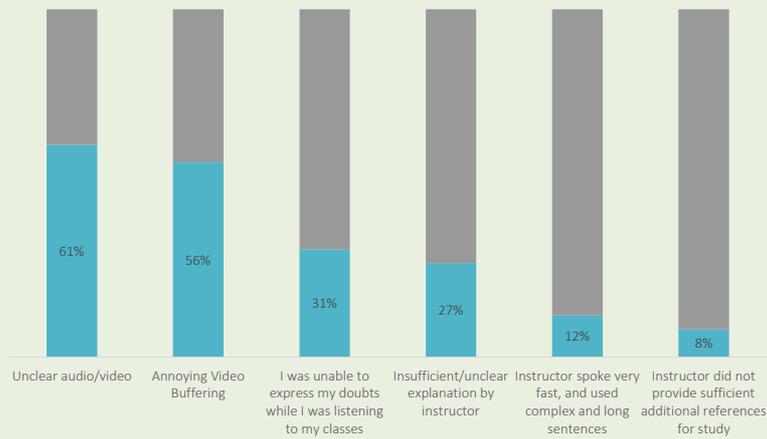

| Unclear audio/video | Annoying Video Buffering | I was unable to express my doubts while I was listening to my classes | Insufficient/unclear explanation by instructor | Instructor spoke very fast, and used complex and long sentences | Instructor did not provide sufficient additional references for study |
|---|---|---|---|---|---|
| 61% | 56% | 31% | 27% | 12% | 8% |

## #Hours - Asynchronous Mode

**Observations:**
About 45% of the students viewed recorded lectures for less than 5 hours.

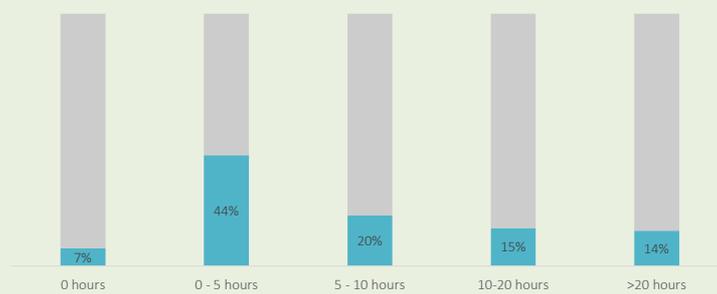

| 0 hours | 0 - 5 hours | 5 - 10 hours | 10-20 hours | >20 hours |
|---|---|---|---|---|
| 7% | 44% | 20% | 15% | 14% |

## % Classes Viewed w/o Difficulty - Asynchronous Mode

**Observations:**
Only 25% of the students were able to view recorded classes without difficulty.

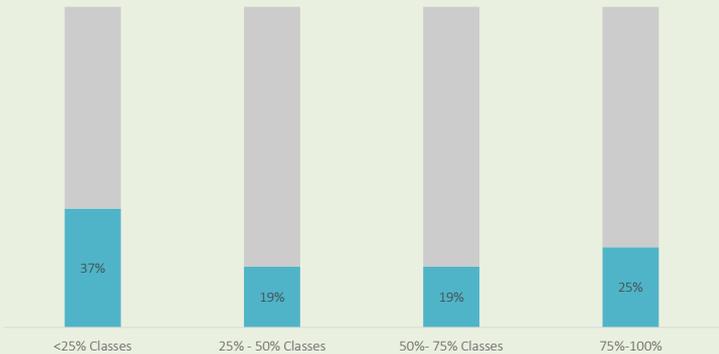

| <25% Classes | 25% - 50% Classes | 50%- 75% Classes | 75%-100% |
|---|---|---|---|
| 37% | 19% | 19% | 25% |

## Nature of Difficulty Faced – Asynchronous Mode

**Observations:**
About 50% of the students were unable to download the material shared with them.

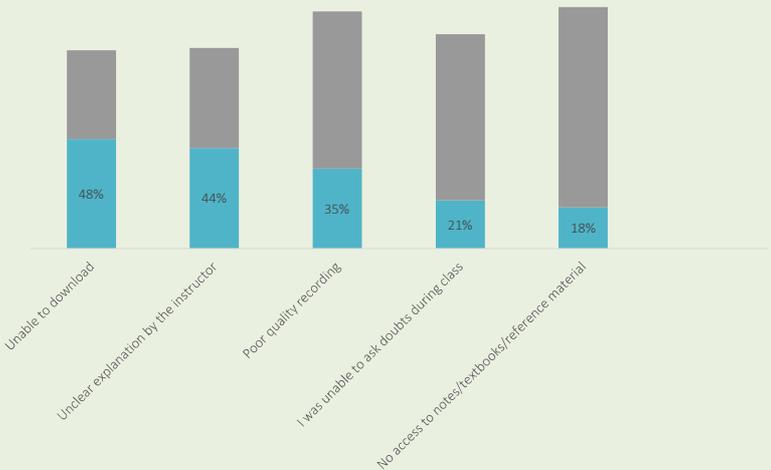

| Unable to download | Unclear explanation by the instructor | Poor quality recording | I was unable to ask doubts during class | No access to notes/textbooks/reference material |
|---|---|---|---|---|
| 48% | 44% | 35% | 21% | 18% |

## Why I Could NOT Clarify Doubts? – Synchronous Mode

**Observations:**
About 50% of the students who responded were not able to ask their doubts due to technological limitation.

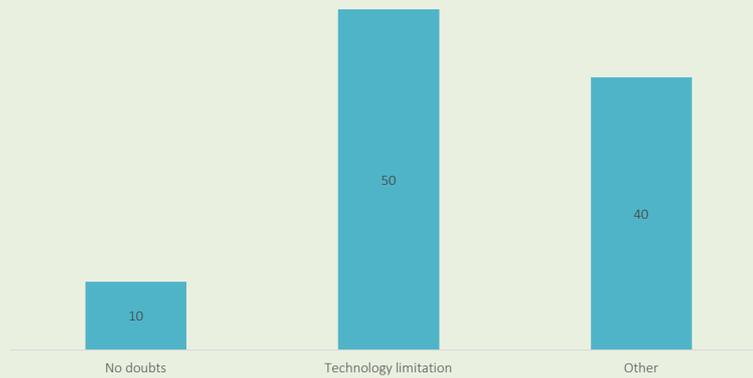

| No doubts | Technology limitation | Other |
|---|---|---|
| 10 | 50 | 40 |

---

## Advantages of Online Education

**Self-Paced Learning**  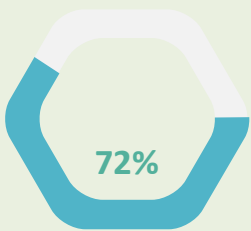 **72%**

**Anywhere-Learning**  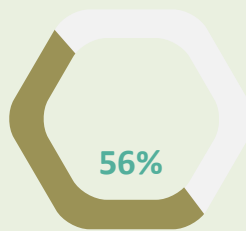 **56%**

**Anytime-Learning**  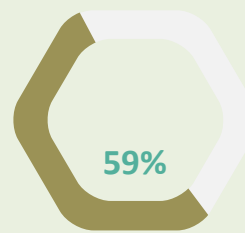 **59%**

**Observations:**
About 72% of the respondents feel that self-paced learning is the biggest advantage of the online education.

## Disadvantages of Online Education

**Lack of personal interaction**

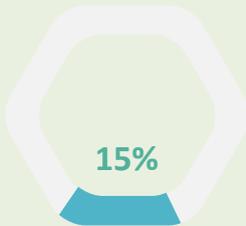

**15%**

**Unable to ask real-time questions and get real-time feedback**

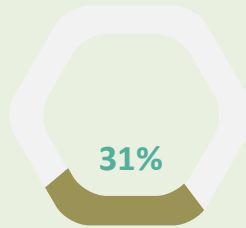

**31%**

**Both**

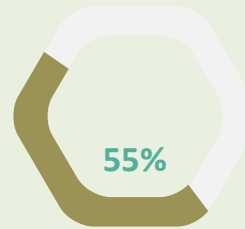

**55%**

**Observations:**

Lack of personal interaction and inability to ask real-time questions and get real-time feedback seems to be the biggest disadvantage of the online education.

---

## ONLINE DELIVERY MODE PREFERENCE

**Only Synchronous Mode**

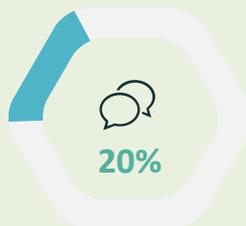

**20%**

**Only Asynchronous Mode**

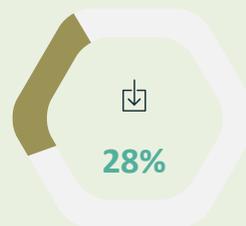

**28%**

**A Combination of Both**

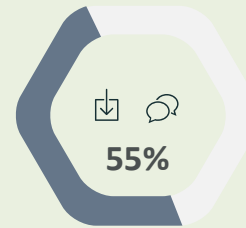

**55%**

**Observations:**

Majority of the students prefer a combination of both synchronous and asynchronous mode of online learning

## Productivity at Home (Compared to Campus)

**Less Productive** 72%

**More Productive** 14%

**Almost the Same** 14%

**Observations:**
Most of the students felt that their productivity was low at home compared to that at their campuses.

---

## How to Improve Online Education?

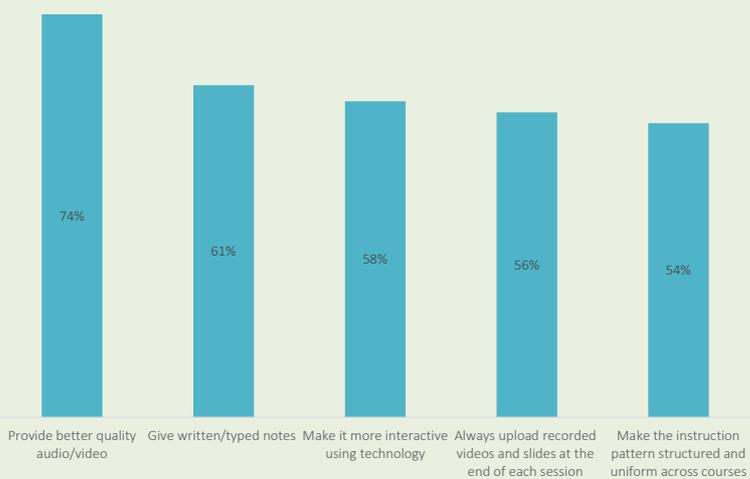

| 74% | 61% | 58% | 56% | 54% |

Provide better quality audio/video | Give written/typed notes | Make it more interactive using technology | Always upload recorded videos and slides at the end of each session | Make the instruction pattern structured and uniform across courses

## How Did You Utilize Time at Home?

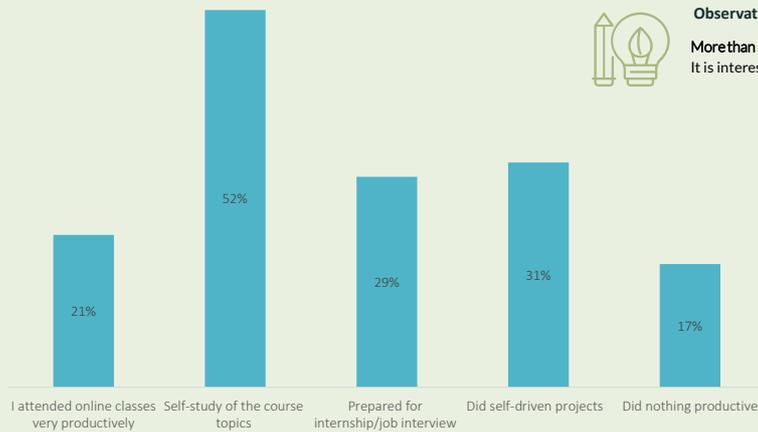

**Observations:**

More than 50% of the students did self-study.
It is interesting to note that about 31% did self-driven projects.

- I attended online classes very productively: 21%
- Self-study of the course topics: 52%
- Prepared for internship/job interview: 29%
- Did self-driven projects: 31%
- Did nothing productive: 17%

## How to Evaluate?

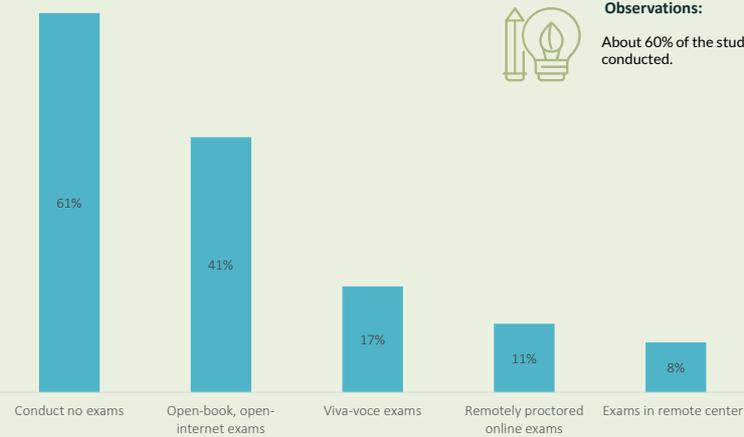

**Observations:**

About 60% of the students felt that no exams should be conducted.

- Conduct no exams: 61%
- Open-book, open-internet exams: 41%
- Viva-voce exams: 17%
- Remotely proctored online exams: 11%
- Exams in remote centers: 8%

## Role of TAs

**Observations:**

Most students did not interact with TAs.

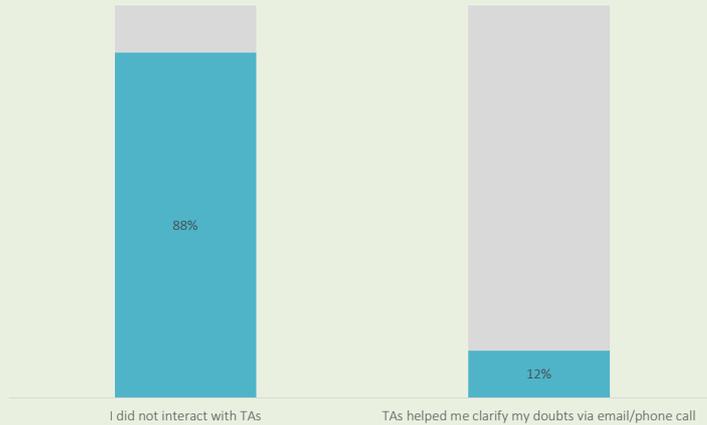

- I did not interact with TAs: 88%
- TAs helped me clarify my doubts via email/phone call: 12%

## Difficulties for Executing the Project

**Observations:**

Only 27% of the students had no difficulty in executing their projects.

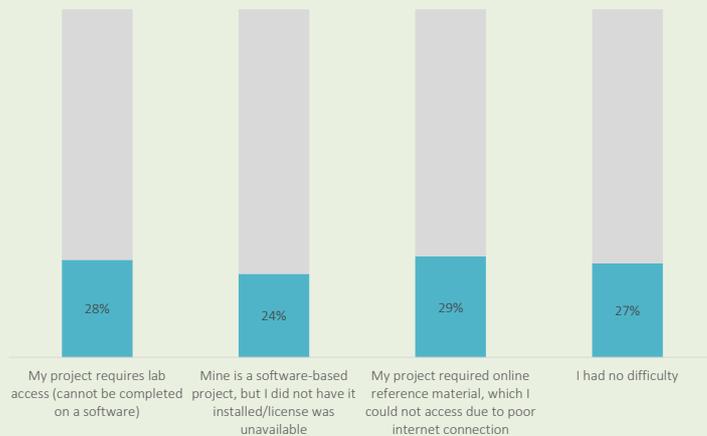

- My project requires lab access (cannot be completed on a software): 28%
- Mine is a software-based project, but I did not have it installed/license was unavailable: 24%
- My project required online reference material, which I could not access due to poor internet connection: 29%
- I had no difficulty: 27%

## Difficulties in Completing Assignments

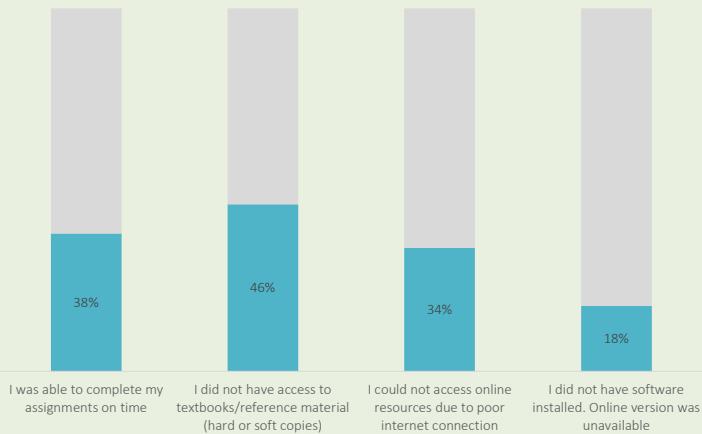

**Observations:**

Only 38% of the students had no difficulty in completing their assignments on time.

Bars:
- I was able to complete my assignments on time: 38%
- I did not have access to textbooks/reference material (hard or soft copies): 46%
- I could not access online resources due to poor internet connection: 34%
- I did not have software installed. Online version was unavailable: 18%

## How Did You Clarify Doubts?

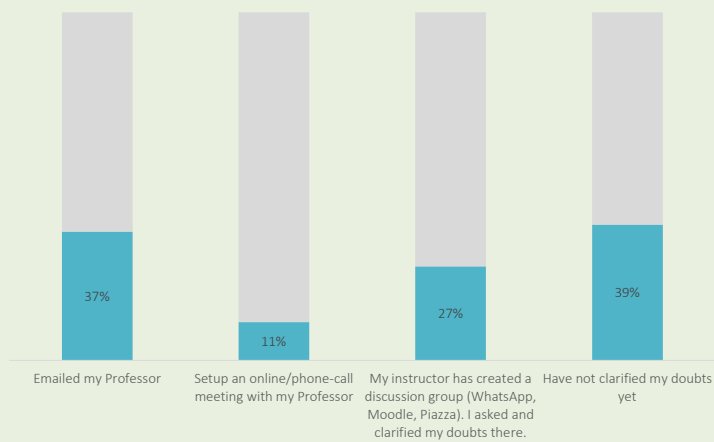

**Observations:**

About 39% of the students have not clarified their doubts yet.

Bars:
- Emailed my Professor: 37%
- Setup an online/phone-call meeting with my Professor: 11%
- My instructor has created a discussion group (WhatsApp, Moodle, Piazza). I asked and clarified my doubts there: 27%
- Have not clarified my doubts yet: 39%

# SECONDARY SCHOOLING

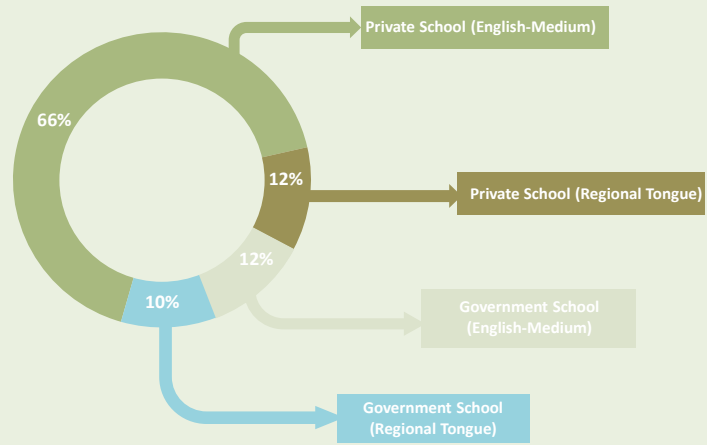

**Inferences:**
Majority of the students completed their secondary education in private, English-medium schools.

- 66% Private School (English-Medium)
- 12% Private School (Regional Tongue)
- 12% Government School (English-Medium)
- 10% Government School (Regional Tongue)

---

# PREFERRED LANGUAGE OF INSTRUCTION

| English | Mother Tongue | Both |
|---------|---------------|------|
| 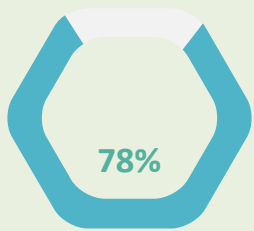 | 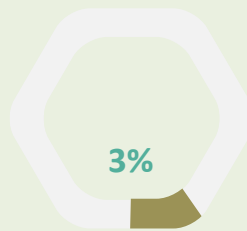 | 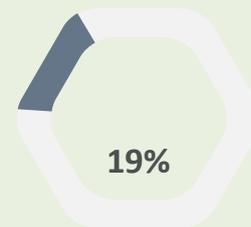 |
| 78% | 3% | 19% |

**Observations:**
Majority of the students prefer instruction in English-Medium. It is important to note that about 22% of the students, like to learn in bi-lingual medium, consisting of ones' mother tongue and English.

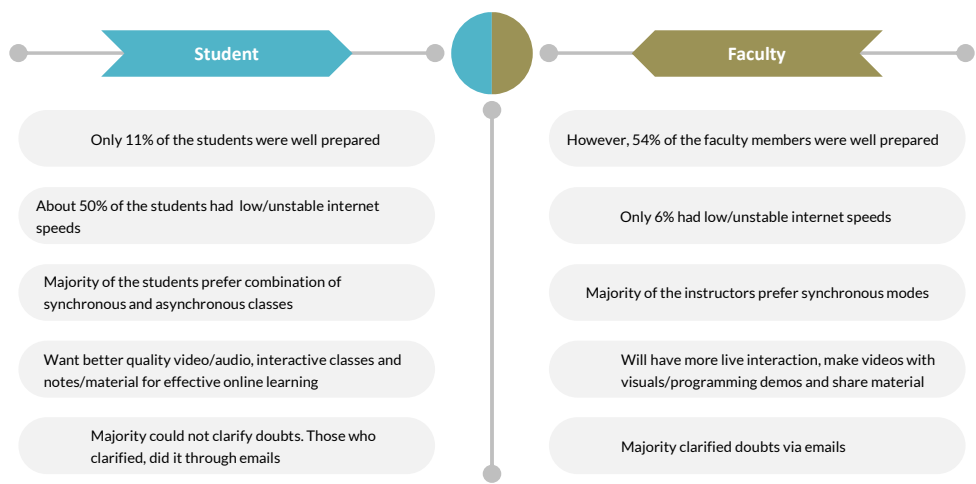

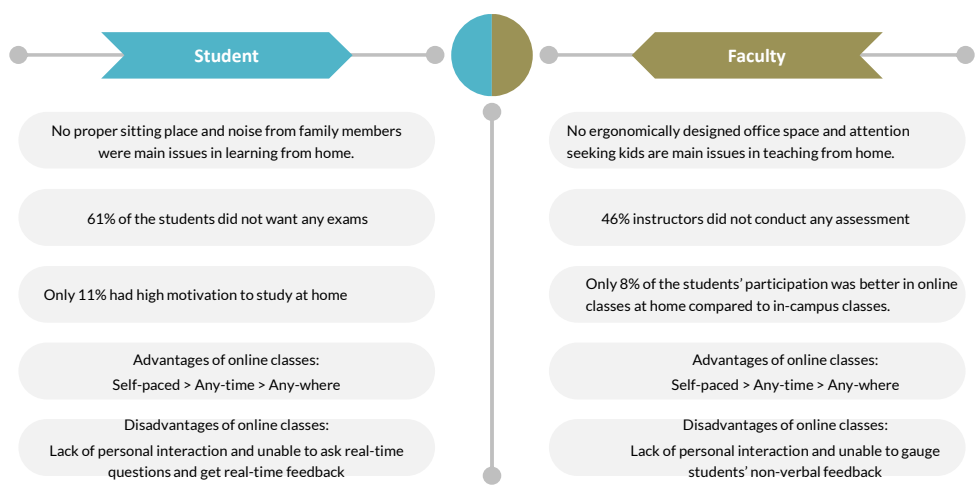